\documentclass[11pt,twoside]{article}
\usepackage{asp2010}
\usepackage{graphicx}
\newcommand{\msun}{{\rm\,M_\odot}}
\newcommand{\kms}{\ifmmode{\,\hbox{km\,s}^{-1}}\else {\rm\,km\,s$^{-1}$}\fi}
\resetcounters

\begin{document}

\title{Counting Dark Sub-halos with Star Stream Gaps }
\author{Raymond G. Carlberg$^1$
\affil{$^1$Department of Astronomy \& Astrophysics, University of Toronto, Toronto, ON, M5S 3H4, Canada. ~{\tt raymond.carlberg@utoronto.ca}}}

\begin{abstract}
The Cold Dark Matter paradigm predicts vast numbers of dark matter sub-halos to be orbiting in galactic halos. The sub-halos are detectable through the gaps they create gaps in stellar streams. The gap-rate is an integral over the density of sub-halos, their mass function, velocity distribution and the dynamical age of the stream.  The rate of {\em visible} gap creation is a function of the width of the stream. The available data for four streams: the NW stream of M31, the Pal~5 stream, the Orphan Stream and the Eastern Banded Structure, are compared to the LCDM predicted relation. We find a remarkably good agreement, although there remains much to be done to improve the quality of the result.  The narrower streams require that there is a total population of order $10^5$ sub-halos above $10^5 \msun$ to create the gaps.  
\end{abstract}

\section{Introduction}

LCDM simulations predict about $10^4$ sub-halos more massive than $10^6\msun$ in a steeply rising mass function that constitutes nearly 10\% of a typical galaxy's dark halo \citep{Aquarius, ViaLactae, Ghalo}.   The large population of sub-halos is not detected in stars, gas, or in any currently detectable annihilation radiation. Logically there are three broad possibilities for the deficiency: dark matter sub-halos never formed with the LCDM predicted abundance, or, the dark matter sub-halos did form and through subsequent evolution were severely depleted, or, the dark matter sub-halos do exist but only a few are populated with visible baryons. Any one of these solutions has significant implications for galaxy formation theory and cosmology.

Here we develop a statistical method to count dark matter sub-halos.   The gravitational field of a sub-halo induces orbital deflections in stellar streams  that are potentially observable as either velocity or positional disturbances \citep{Ibata:02, SGV:08,YJH:11, M31_NW:11}.  In particular, close passages create gaps which are detectable in measurements of the density along a  star stream.  We provide an expression for the average rate at which gaps are created in star streams that are populated with the full LCDM population of dark matter sub-halos. Combining this with the predicted length of the gaps gives a predicted relationship between two observable quantities, the gap creation rate and the width of a star stream.

\section{The Gap-Rate vs Width Relation}

\begin{figure} 
\begin{center} 
\includegraphics[scale=0.19]{den3.eps}
\includegraphics[scale=0.19]{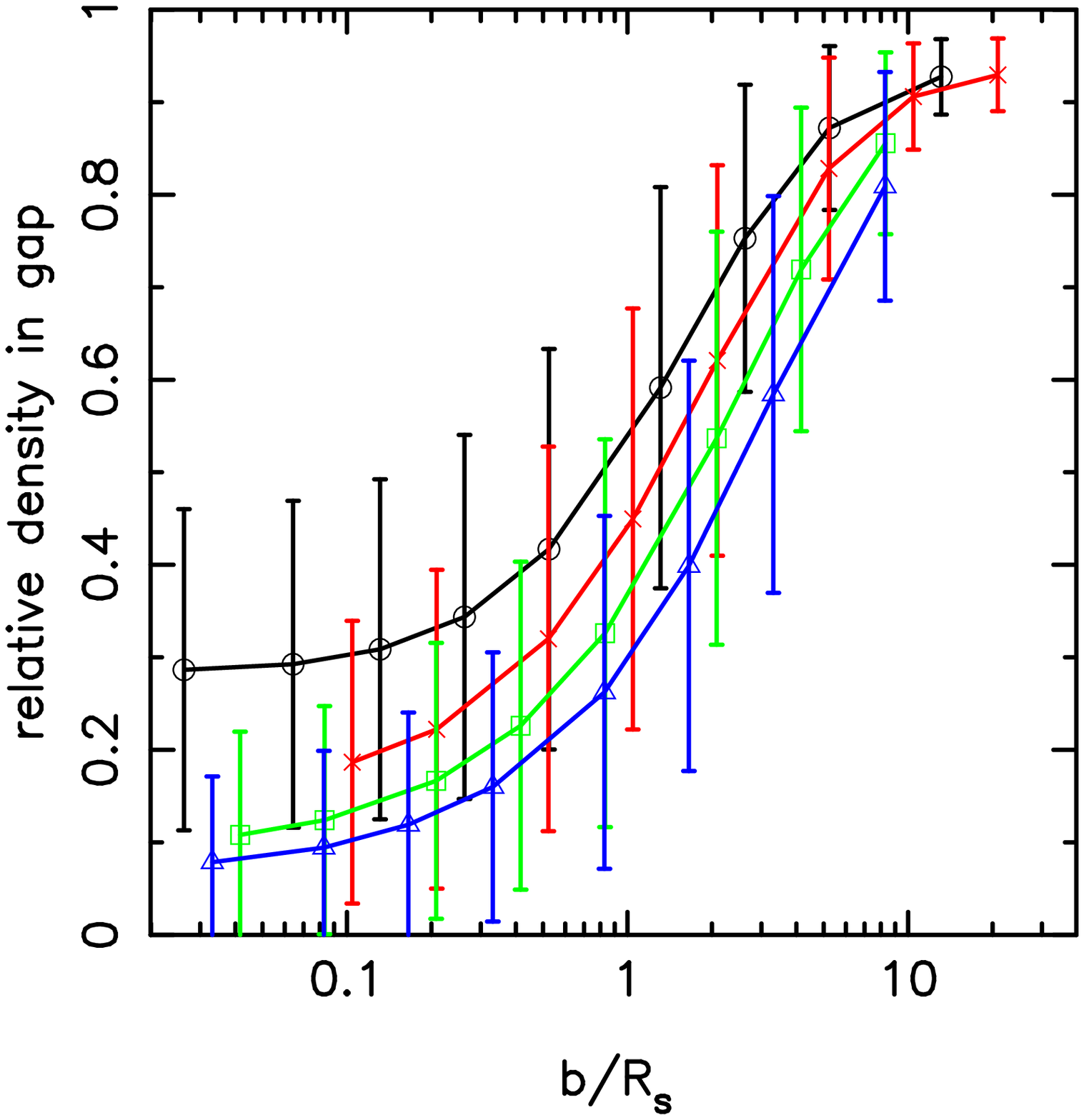}
\includegraphics[scale=0.19]{binzmax3.eps}
\includegraphics[scale=0.19]{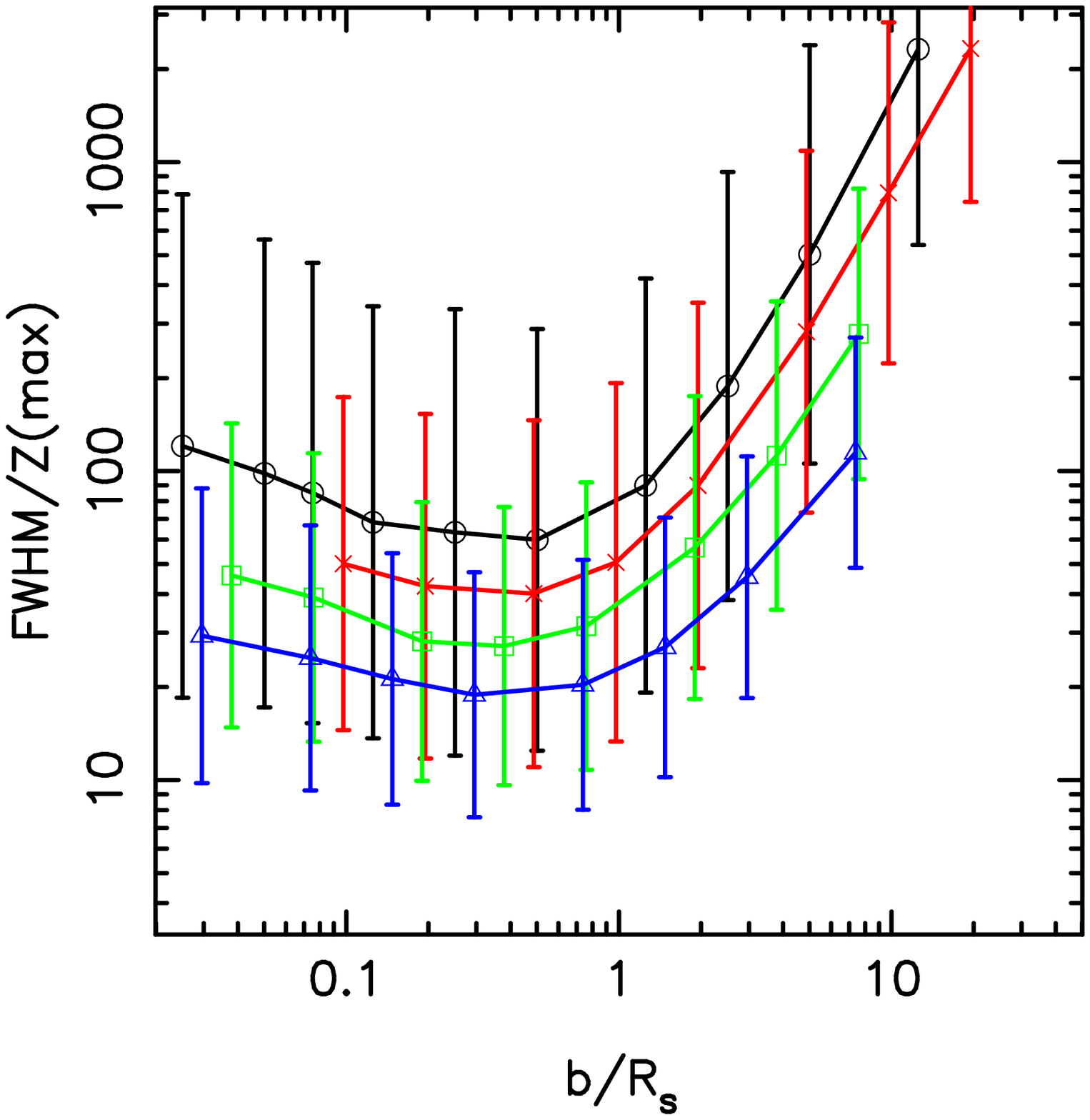}
\end{center}
\caption{ First panel: The density at the location of closest passage, showing decreasing dip depth with increasing impact parameter. Second panel: The depth of the density dip as a function of the impact parameter ratioed to the sub-halo scale radius for masses increasing from $10^6$ to $10^9\msun$. Third panel: The vertical extent of the loop for rings at 30, 60 and 90 kpc with the impact approximation prediction for 60 kpc  is shown as the line. Fourth panel: The ratio of the gap length to the vertical extent, again for  masses increasing from $10^6$ to $10^9\msun$.}
\label{fig_sims}
\end{figure}

The rate per unit length at which sub-halos cross the stream at velocity $v_\perp$ to create gaps is,
\begin{equation}
{\mathcal R}_\cup(r) = \int_M \int_{v_\perp} \int_0^{b_{max}} n(r) N(M) v_\perp f(v_\perp) \pi\, db \, dv_\perp \, dM,
\label{eq_Ratedef}
\end{equation}
where $b_{max}$ is the biggest impact parameter that can create a visible gap.
The various factors in Equation~\ref{eq_Ratedef} are evaluated with numerical simulations. We take the sub-halo mass function $N(M)$ and radial  distribution $n(r)$ from \citet{Aquarius}. We perform a large set of simulations of rings with single halo fly-bys to determine the depth and length of the gap that a sub-halo induces. It is quickly apparent that only encounters where the sub-halo partially passes through the star stream lead to a noticeable density dip. 
We describe the relation as a step function at $b\le \alpha R_s(M)$ where $R_s(M)$ is the scale radius of the sub-halo. We use the \citet{Aquarius} measurement for $n(r)$ (their Fig. 11) which finds that the sub-halo density at 100 kpc is 6 times the mean within the $V_{50}$ volume.   The simulations provide the necessary relations for the maximum effective impact parameter and length of the gap, as shown in Figure~\ref{fig_sims}. 

The width of the stream is set by the velocity dispersion around a mean guiding center motion of the stream. The epicyclic motion of stars around the guiding center means that to a good approximation that any gap narrower than the stream width will be blurred out by the random motion in the stream. Therefore the stream width is the minimum visible gap size and we can eliminate $\hat{M}$ from our equations through $w=l(\hat{M})$.  Putting all these factors together gives gap creation rate as a function of the width of stream, 
\begin{equation}
{\mathcal R}_\cup(w,r) = 0.059 \left( {n(r)/n_0\over{6}}\right) \left( {r\over{\rm 100\, kpc}}\right)^{0.55}
w^{-0.85}\, {\rm kpc}^{-1} {\rm Gyr}^{-1}, 
\label{eq_w_rate}
\end{equation}
for $w$ in kpc. Equation~\ref{eq_w_rate} provides a straightforward testable prediction  of the CDM halo substructure model, effectively counting sub-halos.

\begin{table}
\caption{Gap Statistics and Widths of Star Streams }
\begin{tabular}{|r|r|r|r|r|r|r|l|}
\hline
Stream & Gaps  & Length & Width   & Age & $R_{GC}$ & $n/n_0$&Reference\\
 & \# & [kpc] & [kpc] & [Gyr]  &[kpc] &  & \\
 \hline
M31-NW 	& 12 	& 200 	& 5.0 		& 10	& 100	&6	 &\citet{M31_NW:11} \\
Pal~5		& 5		& 6.5	& 0.11	& 7		& 19	&24  & \citet{GD:Pal5} \\
EBS			& 8		& 4.7	& 0.17	& 7		& 15 	&30	 & \citet{Grillmair:11} \\
Orphan 		& 2 		& 30 	&1.0 	& 3.9	& 30	 &20 & \citet{Newberg:11} \\
\hline
\end{tabular}
\end{table}

\begin{figure} 
\begin{center} 
\includegraphics[scale=0.6]{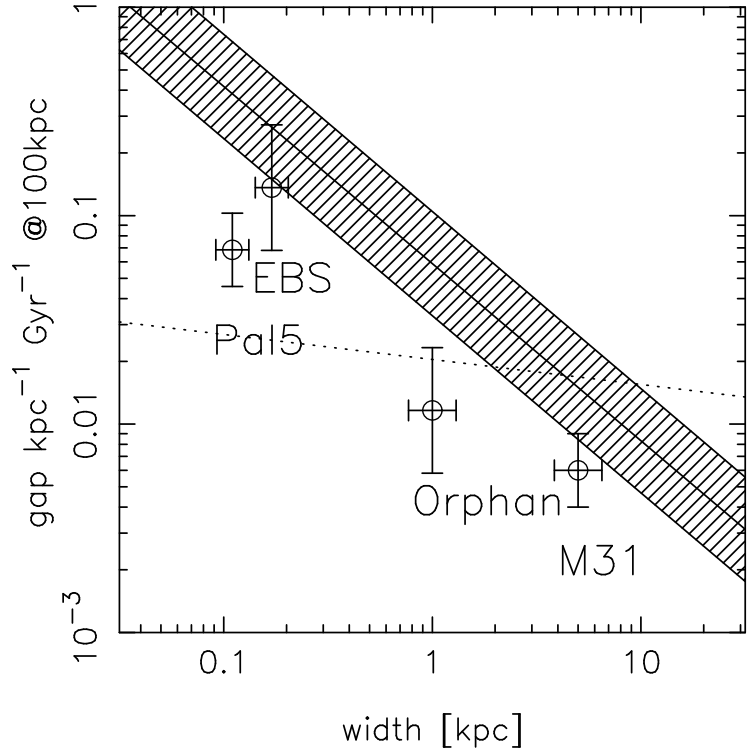}
\end{center}
\caption{The estimated gap rate vs stream width relation for M31 NW, Pal~5, the EBS and the CDM halo prediction.  All data have been normalized to 100 kpc.  Predictions for an arbitrary alternative mass functions, $N(M)\propto M^{-1.6}$ normalized to have the same number, 33, halos above $10^9\msun$ as the standard distribution, is shown with a dotted line. The predicted relation has a width due to the width of the fitted relationships, as well as a halo-to-halo variance. }
\label{fig_gaprate}
\end{figure}

\section{Comparison of Predictions and Observations}

The observed stream gap rate and width estimates are summarized in Table~1. The results of this paper are largely encapsulated in Figure~\ref{fig_gaprate}, which plots the measured and predicted gap rates against the widths of the streams. The data have been scaled for plotting to a galacto-centric distance of 100 kpc using $n(r)/n_0$ and $r$ in Equation~\ref{eq_w_rate}. Comparison of the CDM based prediction of the gap rate-width relation with the published density data for four streams shows generally  good agreement within the fairly generous measurement errors. The result is a statistical argument that the vast predicted population of sub-halos is indeed present in the halos of galaxies like M31 and the Milky Way. 

\noindent
{\bf Question:} (Michael Rich) Doesn't the existence of wide binaries limit the number of sub-halos that can be present in a galactic halo? {\bf Reply:}  Probably not. The distribution of wide binaries is said to be an inverse power (Opik's) law (Longhitano \& Binggeli 2010) to the galactic tidal limit at 1pc, with no features. The origin of this distribution is unclear. Penarrubia etal (2010) have done some simulations of the survivability of wide binaries in the very low velocity dispersion environment of a dwarf galaxy, finding that wide binaries should be eroded to a roughly inverse square law beyond $10^4$ to $10^5$ au (i.e. nearly a parsec) depending on sub-halo fraction.  Therefore it appears that the effect will cut in about where the distribution is depleted anyway, so at present there is no clear problem from wide-binaries. Carr \& Sakellariadou considered this issue in a more general context and found the very hard potentials of globular clusters to be more of a problem.

\noindent
{\bf Question:} (Joss Bland-Hawthorn) Could the stream clumps be due to ``bead" instabilities? {\bf Reply:} This is a complex and controversial topic. Kupper, Heggie and others have pointed out that the first one or two clumps close to a globular cluster progenitor are likely the result of epicyclic pileup. Further down the stream our current full n-body simulations find no instabilities, but much depends on the dark matter content.

\acknowledgements The author thanks NSERC and the Canadian Institute for Advanced Research for support.

\bibliography{carlberg}

\begin{thebibliography}{}
\expandafter\ifx\csname natexlab\endcsname\relax\def\natexlab#1{#1}\fi
\expandafter\ifx\csname url\endcsname\relax
  \def\url#1{\texttt{#1}}\fi
\expandafter\ifx\csname urlprefix\endcsname\relax\def\urlprefix{URL }\fi
\providecommand{\eprint}[2][]{\url{#2}}

\bibitem[{{Carlberg} et~al.(2011){Carlberg}, {Richer}, {McConnachie}, {Irwin},
  {Ibata}, {Dotter}, {Chapman}, {Fardal}, {Ferguson}, {Lewis}, {Navarro},
  {Puzia}, \& {Valls-Gabaud}}]{M31_NW:11}
{Carlberg}, R.~G., {Richer}, H.~B., {McConnachie}, A.~W., {Irwin}, M., {Ibata},
  R.~A., {Dotter}, A.~L., {Chapman}, S., {Fardal}, M., {Ferguson}, A.~M.~N.,
  {Lewis}, G.~F., {Navarro}, J.~F., {Puzia}, T.~H., \& {Valls-Gabaud}, D. 2011,
  \apj, 731, 124. \eprint{1102.3501}

\bibitem[{{Diemand} et~al.(2007){Diemand}, {Kuhlen}, \& {Madau}}]{ViaLactae}
{Diemand}, J., {Kuhlen}, M., \& {Madau}, P. 2007, \apj, 667, 859.
  \eprint{arXiv:astro-ph/0703337}

\bibitem[{{Grillmair}(2011)}]{Grillmair:11}
{Grillmair}, C.~J. 2011, \apj, 738, 98. \eprint{1107.5044}

\bibitem[{{Grillmair} \& {Dionatos}(2006)}]{GD:Pal5}
{Grillmair}, C.~J., \& {Dionatos}, O. 2006, \apjl, 641, L37.
  \eprint{arXiv:astro-ph/0603062}

\bibitem[{{Ibata} et~al.(2002){Ibata}, {Lewis}, {Irwin}, \& {Quinn}}]{Ibata:02}
{Ibata}, R.~A., {Lewis}, G.~F., {Irwin}, M.~J., \& {Quinn}, T. 2002, \mnras,
  332, 915. \eprint{arXiv:astro-ph/0110690}

\bibitem[{{Newberg} et~al.(2010){Newberg}, {Willett}, {Yanny}, \&
  {Xu}}]{Newberg:11}
{Newberg}, H.~J., {Willett}, B.~A., {Yanny}, B., \& {Xu}, Y. 2010, \apj, 711,
  32. \eprint{1001.0576}

\bibitem[{{Siegal-Gaskins} \& {Valluri}(2008)}]{SGV:08}
{Siegal-Gaskins}, J.~M., \& {Valluri}, M. 2008, \apj, 681, 40.
  \eprint{0710.0385}

\bibitem[{{Springel} et~al.(2008){Springel}, {Wang}, {Vogelsberger}, {Ludlow},
  {Jenkins}, {Helmi}, {Navarro}, {Frenk}, \& {White}}]{Aquarius}
{Springel}, V., {Wang}, J., {Vogelsberger}, M., {Ludlow}, A., {Jenkins}, A.,
  {Helmi}, A., {Navarro}, J.~F., {Frenk}, C.~S., \& {White}, S.~D.~M. 2008,
  \mnras, 391, 1685. \eprint{0809.0898}

\bibitem[{{Stadel} et~al.(2009){Stadel}, {Potter}, {Moore}, {Diemand}, {Madau},
  {Zemp}, {Kuhlen}, \& {Quilis}}]{Ghalo}
{Stadel}, J., {Potter}, D., {Moore}, B., {Diemand}, J., {Madau}, P., {Zemp},
  M., {Kuhlen}, M., \& {Quilis}, V. 2009, \mnras, 398, L21. \eprint{0808.2981}

\bibitem[{{Yoon} et~al.(2011){Yoon}, {Johnston}, \& {Hogg}}]{YJH:11}
{Yoon}, J.~H., {Johnston}, K.~V., \& {Hogg}, D.~W. 2011, \apj, 731, 58.
  \eprint{1012.2884}

\end{thebibliography}

\end{document}